\documentclass[letter]{aa}  
\usepackage{graphicx}
\usepackage{txfonts}

\def\kelt{KELT-9\,b}

\def\kms{km\,s$^{-1}$}
\def\h2o{H$_2$O}

\def\pbeta{Pa$\beta$}

\begin{document} 

\title{Detection of Paschen\,$\beta$ absorption in the atmosphere of \kelt}
\subtitle{A new window into the atmospheres of ultra-hot Jupiters}
\titlerunning{Detection of Paschen\,$\beta$ absorption in the atmosphere of \kelt}
\author{
A.~S\'anchez-L\'opez\inst{1},
L.~Lin\inst{1},
I.~A.~G.~Snellen\inst{1},
N.~Casasayas-Barris\inst{1},
A.~Garc\'ia\,Mu\~noz\inst{2},
M.~Lamp{\'o}n\inst{3}, 
M.~L\'opez-Puertas\inst{3}
}

\institute{Leiden Observatory, Leiden University, Postbus 9513, 2300 RA, Leiden, The Netherlands\\
\email{alexsl@strw.leidenuniv.nl}
\and
Universit\'e Paris-Saclay, Universit\'e Paris Cit\'e, CEA, CNRS, AIM, 91191, Gif-sur-Yvette, France
\and
Instituto de Astrof{\'i}sica de Andaluc{\'i}a (IAA-CSIC), Glorieta de la Astronom{\'i}a s/n, 18008 Granada, Spain
}

\authorrunning{A.~S\'anchez-L\'opez et al.}
\date{Received XX / Accepted YY}

\abstract
{Hydrogen and helium transmission signals trace the upper atmospheres of hot gas-giant exoplanets, where the incoming stellar extreme ultraviolet and X-ray fluxes are deposited. Further, for the hottest stars, the near-ultraviolet excitation of hydrogen in the Balmer continuum may play a dominant role in controlling the atmospheric temperature and driving photoevaporation.\\
\kelt\ is the archetypal example of such an environment as it is the hottest gas-giant exoplanet known to date (T$_{\textnormal{eq}}$\,$\sim$\,4500\,K) and orbits an A0V-type star. Studies of the upper atmosphere and escaping gas of this ultra-hot Jupiter have targeted the absorption in the Balmer series of hydrogen (n$_1$\,=\,2\,$\rightarrow$\,n$_2$\,$>$\,2). Unfortunately, the lowermost metastable helium state that causes the triplet absorption at 1083\,$\AA$ is not sufficiently populated for detection. This is due to the low extreme-ultraviolet and X-ray fluxes from the host star, and to its high near-ultraviolet flux, which depopulates this metastable state.
Here, we present evidence of hydrogen absorption in the Paschen series in the transmission spectrum of \kelt\ observed with the high-resolution spectrograph CARMENES. Specifically, we focus on the strongest line covered by its near-infrared channel, \pbeta\ at 12821.6\,$\AA$ (n$_1$\,=\,3\,$\rightarrow$\,n$_2$\,=\,5).  The observed absorption shows a contrast of (0.53\,$^{+0.12}_{-0.13}$)\%, a blueshift of $-$14.8\,$^{+3.5}_{-3.2}$\,\kms, and a full width at half maximum of 31.9$^{+11.8}_{-8.3}$\,\kms. The observed blueshift in the absorption feature could be explained by day-to-night circulation within the gravitationally bound atmosphere or, alternatively, by \pbeta\ absorption originating in a tail of escaping gas moving toward the observer as a result of extreme atmospheric evaporation. This detection opens a new window for investigating the atmospheres of ultra-hot Jupiters, providing additional constraints of their temperature structure, mass-loss rates, and dynamics for future modeling of their scorching atmospheres. 
}

\maketitle

%

\section{Introduction}
\label{Intro}

Ultra-hot Jupiters (UHJs) are highly irradiated gas-giant exoplanets, orbiting extremely close to their host starts, in expected tidally locked configurations \citep[see, e.g.,][]{parmentier2018thermal, bell2018increased, arcangeli2018h}. At their extremely short orbital distances, their daysides receive permanent stellar irradiation, elevating their temperatures well above 2200\,K \citep[see, e.g.,][]{tan2019atmospheric}. Such extreme environments result in complex atmospheric dynamics and potentially strong temperature and chemical gradients across their atmospheres \citep{seidel2019hot, hoeijmakers2019spectral, seidel2021into, ehrenreich2020nightside, kesseli2021confirmation, wardenier2021decomposing, cont2021detection, kesseli2022atomic, savel2022no, cont2022silicon, landman2021, sanchez2022searching}. 

\begin{figure*}[htb!]
\centering
\includegraphics[angle=0, width=1.5\columnwidth]{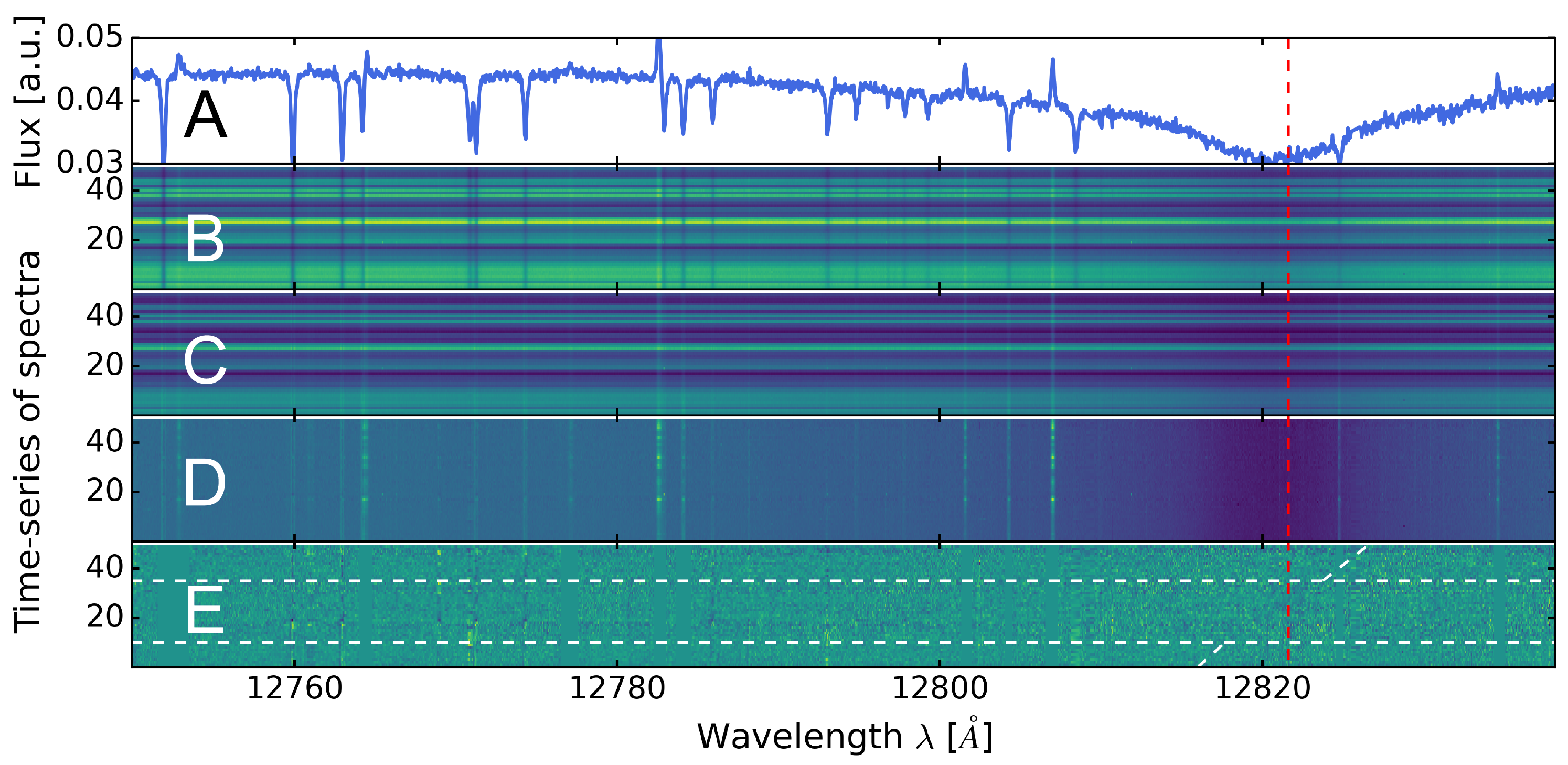}
\caption{Observed flux and spectral matrices through the different analysis steps performed in the rest frame of the Earth. Panel A shows the observed stellar flux around mid-transit as a function of wavelength, and panel B shows the original spectral matrix observed (flux as a function of wavelength and time). A telluric-corrected version of this matrix is shown in panel C. The normalized and telluric-corrected matrix is depicted in panel D, and the residual spectral matrix obtained after removing the stellar contribution and masking noisy columns is shown in panel E. 
The horizontal lines mark the beginning and end of the planet's transit. The tilted lines mark the expected trail of a potential \pbeta\ signal from the planet. The vertical line spanning all panels marks the vacuum wavelength of the \pbeta\ transition. A zoomed-in version of panel E around the expected trail of the planet's \pbeta\ line is shown in Fig.\,\ref{fig:results_fig}B.}
\label{fig:spectral_matrices}
\end{figure*}

Recent studies have shown that low- and high-resolution transmission spectra can probe the escaping material of hot gas giants, providing information about their atmospheric dynamics and mass-loss rates \citep{linsky2010observations, ehrenreich2011mass, yan2018extended, munoz2019rapid, lampon2020modelling, wyttenbach2020mass, lampon2021modelling, casasayas2021carmenes}. Specifically, the Balmer series of hydrogen has been extensively used along with the He I triplet at 10833\,$\AA$ to explore the upper atmosphere of exoplanets \citep{spake2018helium, yan2018extended, casasayas2018and,casasayas2019atmospheric,casasayas2021carmenes, nortmann2018ground, allart2018spectrally,salz2018detection, mansfield2018detection, kreidberg2018non, allart2019high, alonso2019he, kirk2020confirmation, palle2020he, ninan2020evidence, wyttenbach2020mass, dos2020search, palle2020he, vissapragada2020constraints, fossati2021non, krishnamurthy2021nondetection, czesla2022halpha, orell2022tentative, zhang2022escaping}. This has allowed us to better understand planetary evolution and diversity, which are strongly linked to hydrodynamic atmospheric mass loss. That is, the atmospheric heating and a potentially catastrophic outflow at supersonic velocities are triggered by energy deposition from stellar photons, especially by extreme ultraviolet and extreme X-ray fluxes \citep[see, e.g.,][]{munoz2007physical, munoz2019rapid, salz2016simulating, salz2016energy, lampon2021evidence}. In the case of the hottest host stars, the near-ultraviolet (NUV) energy deposition at high altitudes is expected to occur primarily in the hydrogen Balmer continuum \citep[n\,=\,$2\rightarrow\infty$; $\lambda$\,<\,364.6\,nm; see][]{munoz2019rapid}, with the NUV flux being more important than the extreme-ultraviolet radiation in determining the planet's mass loss. This could partially explain the evaporation desert \citep{lopez2012thermal, fulton2017california, jin2018compositional}, at least around hot stars, as the predicted mass-loss rate can be up to two orders of magnitude higher than when only considering extreme ultraviolet energy deposition \citep{munoz2019rapid}.

An extreme and particularly interesting UHJ is \kelt\ \citep{gaudi2017giant}, the hottest gas-giant exoplanet known to date, with an equilibrium temperature of $\sim$4500\,K \citep{mansfield2020evidence}. \kelt\ presents an equilibrium temperature comparable to the effective temperature of many cool stars, which suggests both types of objects may have similar atmospheric features \citep[see, e.g.,][]{parmentier2018thermal, arcangeli2018h, bell2018increased}. A very extended hydrogen envelope (close to the $\sim$1.9\,R$_p$ Roche lobe) was detected by \citet{yan2018extended}, studying the Balmer H$\alpha$ line (n$_1$\,=\,2\,$\rightarrow$\,n$_2$\,=\,3; 6565\,$\AA$), in a transit of this UHJ. This indicates that its atmosphere is evaporating at a notable rate of $\sim$10$^{12}$\,gs$^{-1}$. This detection was confirmed and additional absorption lines from the Balmer series were reported by \citet{wyttenbach2020mass}, who performed a comprehensive study of the thermospheric temperature (13200$^{+800}_{-720}$\,K) and mass-loss rate (10$^{12.8\,\pm\,0.3}$\,gs$^{-1}$). Interestingly, they found that the escape regime of \kelt\ could be ``Balmer-driven,'' as proposed by \citet{munoz2019rapid}.

Due to the lack of a corona and the strong NUV flux of its early-A0V host star KELT-9 \citep[HD 195689, T$_{\textnormal{eff}}$\,$\sim$10170\,K; see, e.g.,][]{gaudi2017giant}, no absorption from the metastable He triplet has been observed in the atmosphere of this planet \citep{nortmann2018ground}. That is, even if radiative recombination and production of the He(2$^3$S) state occurs, its lifetime will be significantly shortened by the stellar NUV flux, which rapidly depopulates this level. As a result, this closes one of the main windows for probing the outer atmospheric layers of this planet. In contrast, the atmosphere of \kelt\ has been surveyed with multiple instruments, yielding a plethora of atmospheric neutral and ionized species \citep{hoeijmakers2018atomic, hoeijmakers2019spectral, cauley2019atmospheric, yan2019ionized, turner2020detection, pino2020neutral, borsa2021high}. Interestingly, these detections point toward potential deviations from the hydrostatic local thermodynamic equilibrium predictions, as evidenced by significantly stronger absorptions than expected in the upper atmosphere.

Here, we aim at expanding the available repertoire of absorption signals suitable for studying exoplanet atmospheres with high-resolution spectrographs. Specifically, we focus on the Paschen series of hydrogen (n$_1$\,=\,3\,$\rightarrow$\,n$_2$\,>\,3), of which the Paschen\,$\beta$ (\pbeta) line at 12821.6\,$\AA$ (n$_1$\,=\,3\,$\rightarrow$\,n$_2$\,=\,5) is the strongest absorption line covered by the near-infrared channel of the CARMENES\footnote{Calar Alto high-Resolution search for M dwarfs with Exoearths with Near-infrared and optical \'Echelle Spectrographs.} instrument \citep{quirrenbach2016carmenes, quirrenbach2018carmenes}. No detection of this line has been presented in the literature for UHJs, although it has been studied in young accreting companions, where the amplitude variations in the line provide information about planet accretion and the time-dependent planet luminosities, temperatures, and radii \citep[see, e.g.,][]{seifahrt2007near, lavigne2009near, baraffe2009episodic, biller2018handbook}.
A detection could open a new window for exploring the atmospheres of UHJs, thus providing new insights into the atmospheric energy budget and dynamics that can complement previous information.

This Letter is organized as follows. In Sect.\,\ref{obs_analysis} we describe the CARMENES observations of this target and the data analysis procedure we followed to obtain the transmission spectrum of \kelt. In Sect.\,\ref{results} we present the results of the analysis and discuss their possible implications. Finally, in Sect.\,\ref{conclusions}, we summarize the main conclusions of this Letter.

\section{Observations and data analysis}
\label{obs_analysis}

We analyzed CARMENES transit data of \kelt\ obtained on August 6, 2017, and available from the Calar Alto public archive. 
The observations consist of 54 exposures with integration times from 300 to 400\,s. This introduces a small orbital blurring, as the planet's orbital velocity with respect to the Earth changes by $\sim$3.5\,\kms\ on average during exposures (the resolution element is $\sim$3\,\kms). The mean signal-to-noise ratio (S/N) per pixel is $\sim$60 in the spectral order containing the \pbeta\ absorption. We discarded four spectra with low signal (S/N\,<\,35) due to passing clouds. Two additional transits of \kelt\ are available in the public archive, observed on September 21, 2017, and 16 June, 2018, but under significantly worse conditions in terms of the mean S/N around the \pbeta\ line or the transit phase coverage, which did not allow us to recover planet signals. We note that the CARMENES optical data of these events have been previously studied in \citet{yan2018extended}, \citet{yan2019ionized}, and \citet{borsa2021high}.
The data were already reduced with the pipeline {\tt caracal} \citep{zechmeister2014flat, caballero2016carmenes}, which includes bias and flat field correction as well as wavelength calibration. The processed spectra are provided with wavelengths in vacuum and in the rest frame of the Earth. To remove possible contamination from cosmic rays, we performed a 3-sigma clipping, interpolating the flagged values (deviated over 3-sigma with respect to the time-series spectra of that pixel) using their adjacent pixels.

The spectral region around \pbeta\ (see Fig.\,\ref{fig:spectral_matrices}) presents a very small telluric contribution that we corrected with the {\tt molecfit} package \citep{smette2015molecfit, kausch2015molecfit}. In Appendix\,\ref{ap:app} we show the regions included and excluded in the analysis. The telluric-corrected spectra were then cut in a region from 12750\,$\AA$ to 12840\,$\AA$ and normalized by the continuum measured away from the stellar \pbeta\ line. For the next part of the analysis, we followed similar methods as those presented in \citet{yan2018extended}, \citet{nortmann2018ground}, \citet{casasayas2019atmospheric}, and \citet{yan2019ionized}. We shifted the spectra to the stellar rest frame using the parameters from Table\,\ref{table:results_table} and computed a master out-of-transit spectrum (F$_{\textnormal{out}}$), combining all of the telluric-corrected spectra that contain no signal from the planet. All the in-transit spectra were divided by this master out-of-transit spectrum to remove the stellar contribution. 

If present, a signal from the atmosphere of \kelt\ is likely to be buried in the noise of the residual spectra. In order to enhance its detectability, the standard approach consists in shifting all in-transit spectra to the rest frame of the planet and co-adding them (F$_{\textnormal{in}}/$F$_{\textnormal{out}}$) before subtracting one to obtain the final transmission spectrum around the \pbeta\ line. In this step, we used a semi-amplitude of the orbital velocity of the planet (K$_P$) of 234.24\,$\pm$\,0.90\kms, as reported by \citet{hoeijmakers2019spectral}, and the in-transit spectra were weighted by their mean S/N.
In order to explore the multidimensional parameter space, we used the statistical framework {\tt MULTINEST} \citep{feroz2009multinest} by means of the {\tt pyMultinest} package \citep{buchner2014x, buchner2016pymultinest}, which allows parameter estimation and model selection. We modeled possible absorption lines assuming they have a Gaussian profile. The retrieval consists of three parameters, namely, the Gaussian line contrast ($C$), its full width at half maximum (FWHM), and the Doppler shift of the Gaussian line center with respect to the vacuum \pbeta\ wavelength, expressed in velocity units ($\varv_{\textnormal{wind}}$).

Following the methods described and applied in \citet{yan2015centre}, \citet{yan2018extended}, and \citet{casasayas2019atmospheric}, we also modeled the distortion in the stellar \pbeta\ line produced by the Rossiter-McLaughlin effect and by the center-to-limb variation. Nevertheless, we found the magnitude of these effects to be almost negligible (not shown), so the correction was not performed in the final analysis.

\section{Results and discussion} 
\label{results}

\begin{figure*}[htb!]
\centering
\includegraphics[angle=0, width=2\columnwidth]{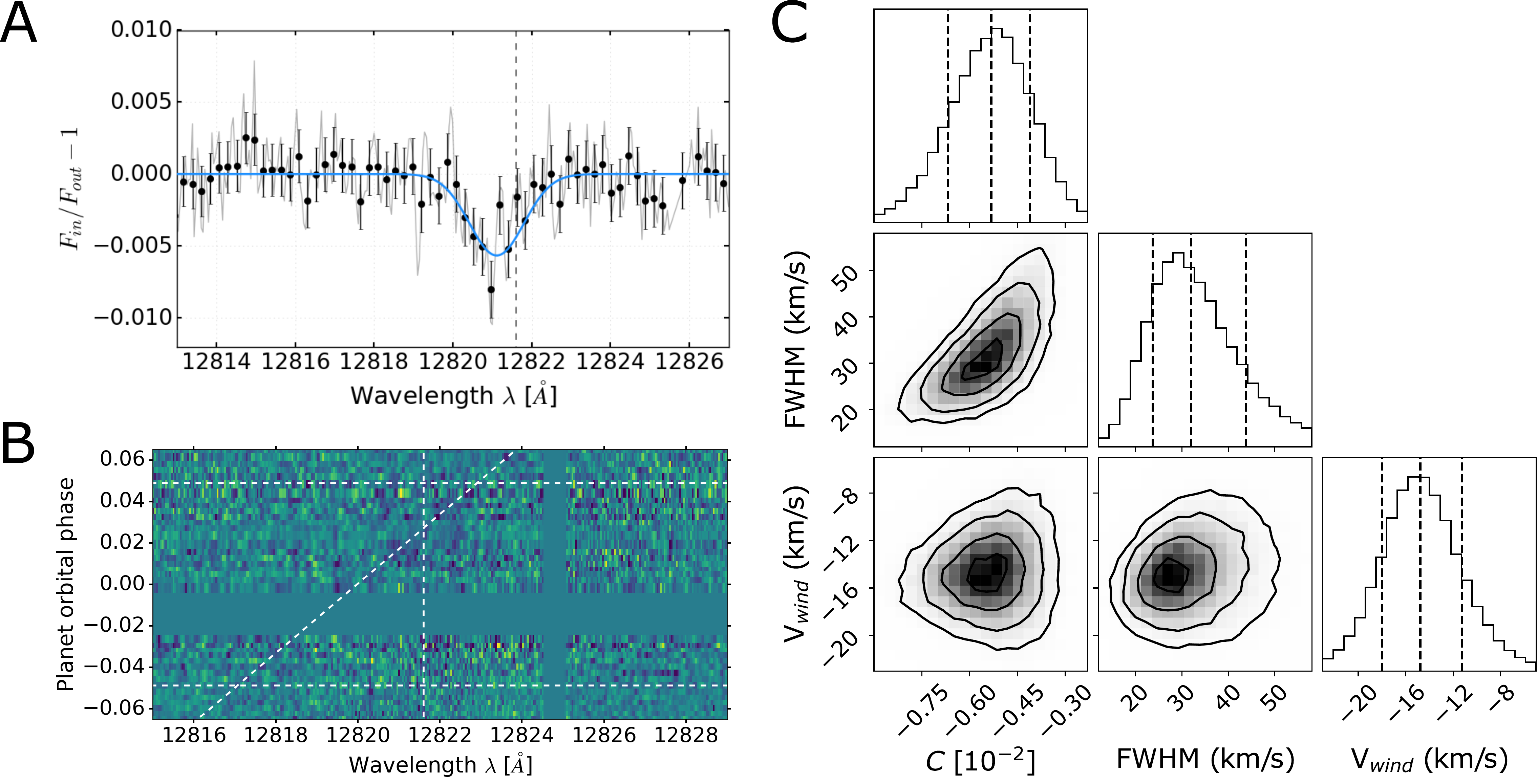}
\caption{Panel A: Transmission spectrum of \kelt\ around the \pbeta\ wavelengths in the rest frame of the exoplanet. The unbinned spectrum is shown in light gray, and the binned spectrum is shown in black, with error bars computed by propagating the instrumental error. The blue curve is the Gaussian best fit to the absorption line we retrieved. The vertical line marks the vacuum wavelength of \pbeta. Panel B: Residual spectral matrix in the rest frame of the Earth obtained after correcting the telluric contribution and dividing out the master out-of-transit spectrum. The horizontal lines mark the starting and ending orbital phases of the planetary transit, the vertical line marks the vacuum wavelength of the \pbeta\ transition, and the tilted line marks the expected trail of potential absorption signals of the exoplanet (combination of orbital planet velocities with respect to the Earth and systemic and barycentric velocities). Noisy pixel columns were masked and excluded from the analyses. No spectra were observed between orbital phases -0.026 and -0.001. Panel C: Probability distributions resulting from the multidimensional analysis of the observed line.}
\label{fig:results_fig}
\end{figure*}

The transmission spectrum shows evidence of hydrogen absorption in the \pbeta\ line in the atmosphere of \kelt\ (see Fig.\,\ref{fig:results_fig} and Table\,\ref{table:results_table}). The retrieved Gaussian fit presents a contrast, $C$, of (0.53\,$^{+0.12}_{-0.13}$)\%, which corresponds to a $\sim$4.2$\sigma$ detection, above the usually assumed 3$\sigma$ threshold. In order to further assess the signal's significance, we compared the $\chi^2$ goodness-of-fit of the retrieved model with that of a flat line (i.e., no signal in the normalized transmission spectrum). We were able to exclude the no-signal hypothesis with a $\sqrt{\Delta\chi^2}\sim$\,6, implying significant evidence for the presence of a real signal in our data. Indeed, the trail originating the observed signal in the transmission spectrum can be seen in Fig.\,\ref{fig:results_fig}B, close to the expected exoplanet velocities during the observations. 

\begin{table}[b!]
\centering
{\tiny
\caption{\label{table:results_table} Parameters used and retrieved in this work.}
\begin{tabular}{ccc} 
\hline
\hline
\noalign{\smallskip}
Parameter & Value & Reference\\
\noalign{\smallskip}
\hline
\noalign{\smallskip}
\noalign{\smallskip}
Physical parameters &  & \citet{gaudi2017giant}\\
\noalign{\smallskip}
R$_{\star}$ & 2.178\,$\pm$\,0.011\,R$_\odot$ & \\ 
\noalign{\smallskip}
M$_{\star}$ & 1.978\,$\pm$\,0.023\,M$_\odot$ & \\ 
\noalign{\smallskip}
T$_{\text{eff}}$ & 10170\,$\pm$\,450\,K & \\ 
\noalign{\smallskip}
R$_{P}$ & 1.783\,$\pm$\,0.009\,R$_J$ & \\ 
\noalign{\smallskip}
M$_{P}$ & 2.44\,$\pm$\,0.70\,M$_J$ & \\
\noalign{\smallskip}
\hline
\noalign{\smallskip}
Transit parameters &  & \citet{gaudi2017giant}\\
\noalign{\smallskip}
T$_0$ & 2457095.68572\,$\pm$\,0.00014\,BJD & \\ 
\noalign{\smallskip}
$P$ & 1.4811235\,$\pm$\,0.0000011\,d & \\ 
\noalign{\smallskip}
$a$ & 0.0346\,$\pm$\,0.001\,au & \\ 
\noalign{\smallskip}
$i$ & 86.79\,$\pm$\,0.25\,\textdegree & \\ 
\noalign{\smallskip}
\hline
\noalign{\smallskip}
Radial velocities &  & \citet{gaudi2017giant}\\
\noalign{\smallskip}
K$_\star$ & 276\,$\pm$\,79\,m\,s$^{-1}$ & \\ 
\noalign{\smallskip}
K$_P$ & 234.24\,$\pm$\,0.90\,\kms\ & \citet{hoeijmakers2019spectral} \\ 
\noalign{\smallskip}
$\varv{_{\text{sys}}}$ & $-$17.74\,$\pm$\,0.11\,\kms\ & \citet{hoeijmakers2019spectral} \\ 
\noalign{\smallskip}
\hline
\noalign{\smallskip}
H$\alpha$\ analysis &  & \citet{yan2018extended}\\
\noalign{\smallskip}
$C$ & (1.15\,$\pm$\,0.05)\% & \\ 
\noalign{\smallskip}
FWHM & 51.2$^{+2.7}_{-2.5}$\,\kms & \\ 
\noalign{\smallskip}
$\varv{_\text{wind}}$ & $-$1.02$^{+0.99}_{-1.00}$\,\kms & \\ 
\noalign{\smallskip}
\hline
\noalign{\smallskip}
\pbeta\ analysis &  &This work\\
\noalign{\smallskip}
$C$ & (0.53\,$^{+0.12}_{-0.13}$)\% & \\
\noalign{\smallskip}
FWHM & 31.9$^{+11.8}_{-8.3}$\,\kms & \\ 
\noalign{\smallskip}
$\varv{_\text{wind}}$ & $-$14.8\,$^{+3.5}_{-3.2}$\,\kms & \\ 
\noalign{\smallskip}
\hline
\end{tabular}
}
\end{table}

The \pbeta\ absorption feature presents a blueshifted $\varv_{\textnormal{wind}}$ of $-$14.8\,$^{+3.5}_{-3.2}$\,\kms\ and a FWHM of 31.9$^{+11.8}_{-8.3}$\,\kms, indicative of hydrogen atoms moving toward the observer and significant broadening of the line profile, respectively (see Table\,\ref{table:results_table}). We note that this FWHM includes significant contributions from rotational and instrumental broadening, $\sim$16 and $\sim$3\,\kms, respectively, in addition to slight orbital blurring. For comparison, the line profile of H$\alpha$ obtained in \citet{yan2018extended}, who also used the optical CARMENES data from August 6, 2017, is slightly broader, with a FWHM of 51.2$^{+2.7}_{-2.5}$\,\kms\ and no indications of day-to-night winds (see Table\,\ref{table:results_table}). 
Thus, the blueshift observed in \pbeta\ is difficult to reconcile with the planet's rest-frame position of H$\alpha$ discussed above, and of other atmospheric signatures detected in transmission by \citet{hoeijmakers2019spectral} and \citet{wyttenbach2020mass}. It is important to caution that the values for the velocity of the KELT-9 planetary system ($v_{\textnormal{sys}}$) found in the literature are somewhat discrepant, being $-$17.74\,$\pm$\,0.11\,\kms\ in \citet{hoeijmakers2019spectral} (the value adopted here) but $-$20.6\,$\pm$\,0.1\,\kms\ in \citet{gaudi2017giant}. This introduces a shift in our retrieved $v_{\textnormal{wind}}$, and we find it to be $-$11.3\,$^{+3.4}_{-3.5}$\,\kms\ when using the value from \citet{gaudi2017giant} \citep[see also the discussion in][]{hoeijmakers2019spectral}. 

Regardless of the different $v_{\textnormal{sys}}$ values, the observed blueshift (at $\sim$\,4.5$\sigma$ for the reference $v_{\textnormal{sys}}=-$17.74\,$\pm$\,0.11\,\kms) still suggests that the atmospheric regions contributing to the observed \pbeta\ absorption are different from those of other excited states of hydrogen. We propose that the H(3)\,$\rightarrow$\,H(5) line observed here is probing lower terminator altitudes ($\sim$\,1.3\,R$_P$) than the H(2)\,$\rightarrow$\,H(3) absorption feature, which is reported to originate at $\sim$\,1.6\,R$_P$ in earlier works \citep{yan2018extended, cauley2019atmospheric, wyttenbach2020mass}. The higher pressures favor the H(3)\,$\rightarrow$\,H(2), H(1) lines remaining optically thick. That is, the smaller population losses of H(3) by spontaneous emission to the lower-energy levels facilitate the buildup of moderately high H(3) densities. In this interpretation, the blueshift we observe could be explained by a day-to-nightside atmospheric flow. 
Global circulation models applied to UHJs suggest that a day-to-nightside wind could be possible in the terminator of \kelt\ at low pressures \citep[see, e.g.,][]{tan2019atmospheric, wong2020exploring}, but the high wind speed we retrieved is not predicted by these models.

An alternative explanation is that \pbeta\ absorption may originate in the atomic hydrogen escaping \kelt, resembling a comet-like tail flowing toward the observer (e.g., along the star-planet axis). This scenario is in principle more unlikely, as the tail densities are expected to be much lower than those of the upper atmosphere. As a result, the H(3) abundance might be too low for detection as the H(3)\,$\rightarrow$\,H(2), H(1) lines are optically thinner. However, \citet{cauley2019atmospheric} propose that small stellar flares arising, for instance, from the star-planet magnetic-field interactions might result in extreme atmospheric expansion events. Although the occurrence of this mechanism is still uncertain for KELT-9, this hypothesis was supported by the P-Cygni-like profiles the authors observed for H$\alpha$, H$\beta$, Mg\,I, and Fe\,II. Such events could result in significantly higher than expected mass-loss rates and densities building up in a relatively narrow tail of escaping gas. 
It is worth recalling that the P-Cygni-like profiles observed in \citet{cauley2019atmospheric} were not reported for H$\alpha$ in \citet{yan2018extended} during the night we analyzed, and we did not detect any \pbeta\ absorption in the post-transit spectra either. However, a recent event at the time of the observations might have significantly expanded the upper atmosphere, potentially enhancing \pbeta\ detectability. In this case, its blueshift might be explained as being due to the interaction with the radiation pressure \citep[see, e.g.,][]{bourrier2016evaporating, spake2018helium, cauley2019atmospheric}, with no significant day-to-nightside circulation being required in the planet.
Under this hypothesis, additional observations could reveal variability in the \pbeta\ absorption depth, which would be expected if the density of the escaping gas changed significantly during an event and shortly after.

We note that a tendency toward stronger, blueshifted winds and smaller FWHMs (compared to H$\alpha$) is hinted at for weaker lines of the Balmer series in \kelt\ in \citet{wyttenbach2020mass} (see their Table\,3). Specifically, the authors reported a blueshift in the H$\delta$ line (H(2)\,$\rightarrow$\,H(6) transition, 4.5$\sigma$ detection) of $-$4.6\,$\pm$\,2.2\,\kms. Unfortunately, the H$\epsilon$ (H(2)\,$\rightarrow$\,H(7) transition) line significance in \citet{wyttenbach2020mass} is just below the detection threshold (2.9$\sigma$), but they reported a blueshift of $-$15.0\,$\pm$\,5.0\,\kms, which is fully consistent with our result. This might indicate that these lines probe similar regions as \pbeta, but more data are needed to either confirm or refute this hypothesis.

Regarding the FWHM of the different hydrogen absorptions, the broad line profile observed for H$\alpha$ in \citet{yan2018extended}, \citet{cauley2019atmospheric}, and \citet{wyttenbach2020mass} cannot be explained by thermal broadening only, as this is about 25\,\kms\ at the thermospheric temperature \citet{wyttenbach2020mass} retrieved ($\sim$\,13200\,K). This is indicative of the H$\alpha$ line being optically thick or forming in a region where the gas is expanding radially at high velocity \citep{munoz2019rapid}. However, considering the uncertainty intervals, thermal broadening could explain the \pbeta\ profile obtained here, with no need for a high radial velocity. Future 3D retrievals of wind patterns in \kelt\ might shed more light onto the widths and Doppler shifts of the observed \pbeta\ and H$\alpha$ lines, which might also have contributions from convective layers, as has been observed for other UHJs \citep[see, e.g.,][]{seidel2021into}.

\section{Conclusions} 
\label{conclusions}
Extremely irradiated exoplanets are perfect astronomical laboratories for searching for spectral features that reveal key information about exoplanet atmospheric temperatures, dynamics, and mass-loss processes. We analyzed CARMENES high-resolution transit observations of the hottest UHJ known to date, \kelt, in a search for hydrogen absorption in the strongest Paschen-series line covered by the instrument's near-infrared channel (\pbeta, 12821.6\,$\AA$). By applying the transmission spectroscopy technique, we observed a \pbeta\ signal around the expected planet velocities with respect to the Earth during the transit. This is the first time this signal has been observed in an UHJ, although it has been widely studied in stellar and young companion atmospheres. As the opacities of the Balmer and Paschen series of lines are expected to be different, H$\alpha$ and \pbeta\ could be probing a significantly different altitude range in the terminator of \kelt. Therefore, our detection expands the available sample of useful lines for exploring the atmosphere of the hottest exoplanet known.  

From the retrieved line's FWHM and blueshift, we propose that this signal could be probing lower atmospheric layers than H$\alpha$, as the higher-density layers greatly favor a longer lifetime of the H(3) population. In this interpretation, the smaller temperature of the lower thermospheric layers and a weaker radial component of the wind would explain the slightly smaller FWHM of \pbeta\ when compared to that of H$\alpha$. In addition, the blueshift observed reveals a strong day-to-night velocity component at the atmospheric layers probed. This wind is not observed with the strongest Balmer lines, although it is hinted at for H$\delta$ and H$\epsilon$ in previous studies.
Alternatively, the origin of \pbeta\ absorption could be in the escaping gas moving toward the observer, accelerated by radiation pressure. Although the small tail densities make it difficult to form a detectable abundance of H(3), the latter might be enhanced if flaring events arising from star-planet magnetic-field interactions proposed in earlier studies are confirmed in future observations.

Therefore, detailed modeling of the H(3) population in this planet is needed so as to accurately constrain the expected atmospheric layers probed with \pbeta. Future observations under better weather conditions will be needed to reproduce this signal and study possible  variability associated with potential flaring events, and in turn provide stronger constrains for future theoretical studies of the complex 3D nature of UHJ atmospheres. Moreover, additional transit observations could be used to search for additional lines in the Paschen series, of which Pa$\gamma$ (H(3)\,$\rightarrow$\,H(6), 10941.1\,$\AA$) is the strongest one in a relatively transparent telluric region.

\begin{acknowledgements}
We acknowledge funding from the European Research Council under the European Union's Horizon 2020 research and innovation program under grant agreement No 694513. IAA-CSIC authors acknowledge financial support from the Agencia Estatal de Investigaci\'on of the Ministerio de Ciencia, Innovaci\'on y Universidades through projects Ref. PID2019-110689RB-I00/AEI/10.13039/501100011033 and the Centre of Excellence ``Severo Ochoa'' award to the Instituto de Astrof\'isica de Andaluc\'ia (SEV-2017-0709). This research has made use of the Spanish Virtual Observatory (http://svo.cab.inta-csic.es) supported by the MINECO/FEDER through grant AyA2017-84089.7.
\end{acknowledgements}

\bibliographystyle{aa} 
\bibliography{ref.bib}

\begin{appendix}

\section{{\tt molecfit} correction around \pbeta}
\label{ap:app}

\begin{figure}[htb!]
\centering
\includegraphics[angle=0, width=1\columnwidth]{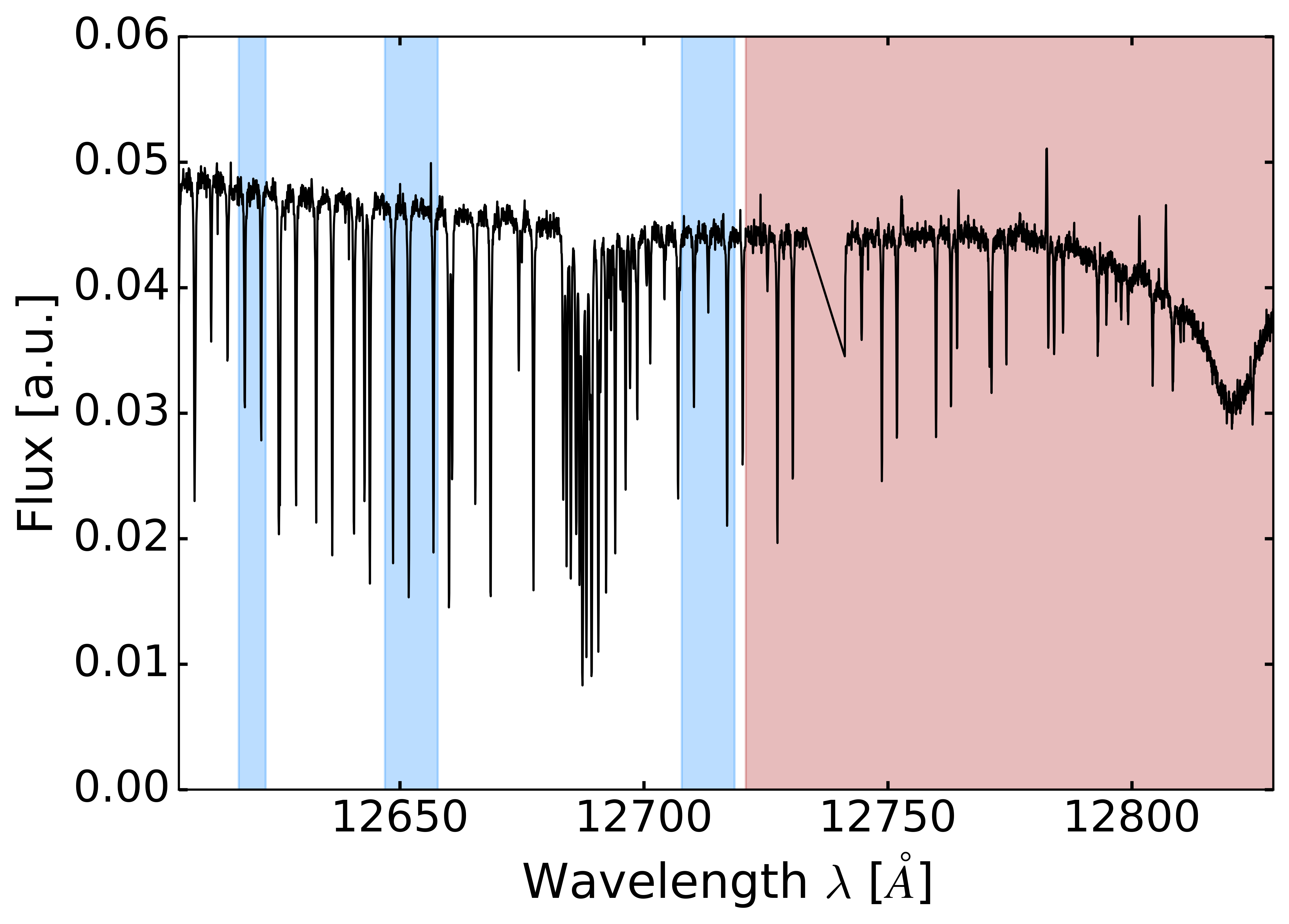}
\includegraphics[angle=0, width=1\columnwidth]{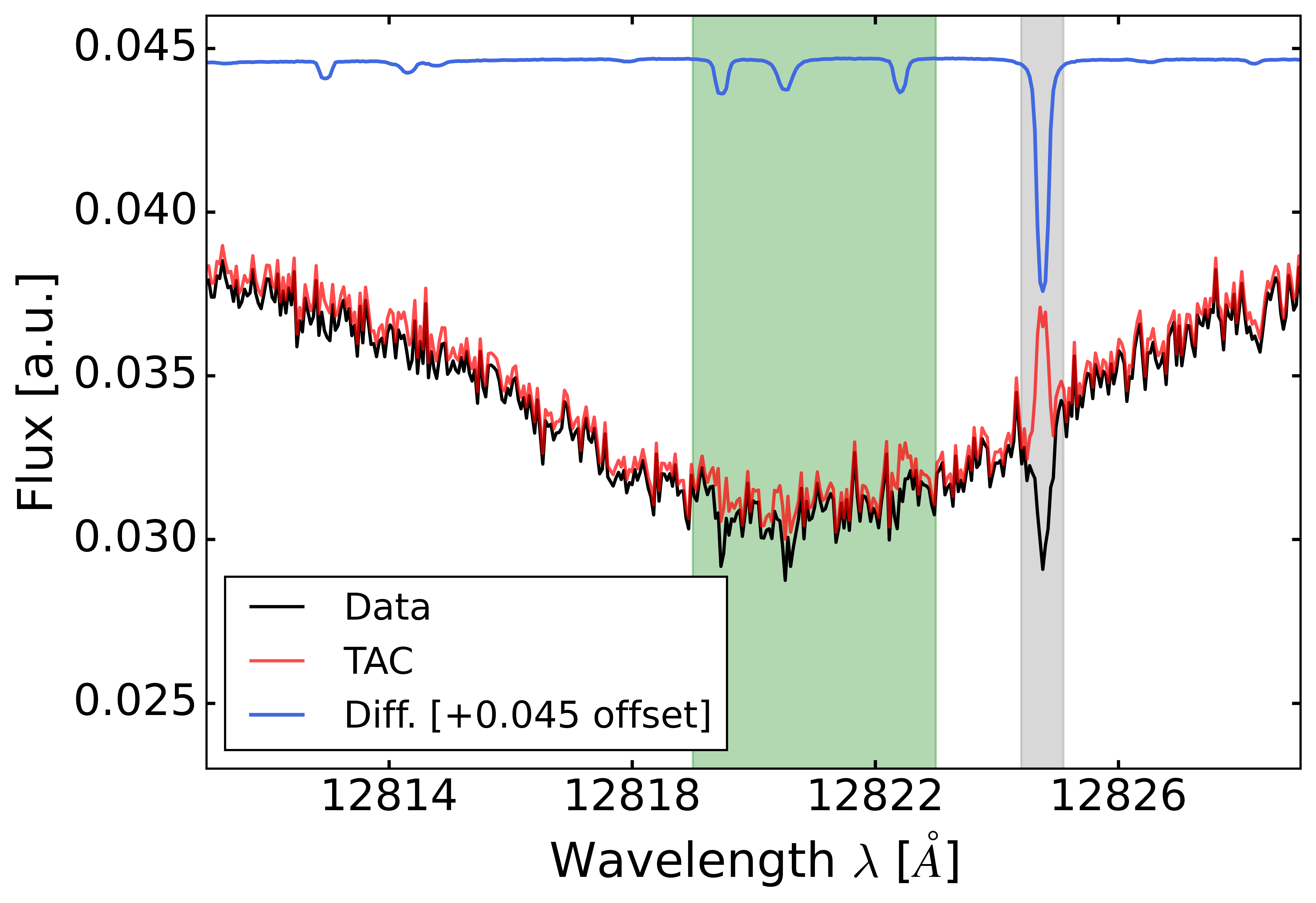}
\caption{Setup and results of the telluric correction with {\tt molecfit}. Upper panel: Flux observed as a function of wavelength at an orbital phase of $-$0.001. The regions included in the {\tt molecfit} analyses contain telluric lines of different depths (with no saturation) and are shaded in blue (spectral regions from 12617.01 to 12622.54\,$\AA$, from 12646.96 to 12657.79\,$\AA$, and from 12707.77 to 12718.60\,$\AA$). The 12720.90$-$12844.15\,$\AA$ region shaded in red contains the stellar \pbeta\ absorption and was excluded from all {\tt molecfit} fits. Bottom panel: Zoomed-in view of the stellar \pbeta\ line. The observed flux from the upper panel is shown in black, the telluric-corrected spectrum (TAC) is shown in red, and the difference between both is shown in blue (with an offset for clarity). The telluric contribution in the spectral region containing the exoplanet \pbeta\ signal (shaded in green) is very small, with negligible residuals after the fit. The spectral region from 12824.4\,$\AA$ to 12825.1\,$\AA$ (shaded in gray), containing a poorly fitted telluric line, was masked for the entire analysis.}
\label{fig:app}
\end{figure}

\end{appendix}

\end{document}